\documentclass[fleqn,usenatbib]{mnras}

\usepackage{newtxtext,newtxmath}

\usepackage[T1]{fontenc}
\usepackage{ae,aecompl}

\usepackage{graphicx}	

\usepackage{amsmath}	
\usepackage{amssymb}	

\title[XMM-Newton Observations of the Plerionic Supernova Remnant CTB~87]{Deciphering the Nature of the Pulsar Wind Nebula CTB~87 with XMM-Newton}

\author[Guest et al.]{B. Guest$^{1}$, S. Safi-Harb$^{1}$, A. MacMaster$^{1}$, R. Kothes$^{2}$
\newauthor
B. Olmi$^{3,4,5}$, E. Amato$^{3,6}$, N. Bucciantini$^{3,6,7}$ \&
Z. Arzoumanian$^{8}$
\\
$^{1}$Department of Physics and Astronomy, University of Manitoba, Winnipeg, MB, R3T 2N2, Canada\\
 $^{2}$ National Research Council Canada, Herzberg Astronomy and Astrophysics, 
Dominion Radio Astrophysical Observatory, \\
P.O. Box 248, Penticton, B.C., V2A 6J9, Canada\\
$^{3}$INAF - Osservatorio Astrofisico di Arcetri, Largo E. Fermi 5,
I-50125 Firenze, Italy\\
$^{4}$Institute of Space Sciences (ICE, CSIC), Campus UAB, Carrer de Magrans s/n, 08193 Barcelona, Spain\\
$^{5}$Institut d'Estudis Espacials de Catalunya (IEEC), 08034 Barcelona, Spain\\
$^{6}$Dipartimento di Fisica e Astronomia, Universit\`a degli Studi di Firenze, Via G. Sansone 1, 
I-50019 Sesto F.~no  (Firenze), Italy\\
$^{7}$INFN - Sezione di Firenze, Via G. Sansone 1, I-50019 Sesto F.~no  (Firenze), Italy\\
$^{8}$ X-ray Astrophysics Laboratory, Code 662, NASA Goddard Space Flight Center, Greenbelt, MD, 20771, USA
}

\date{Accepted XXX. Received YYY; in original form ZZZ}

\pubyear{2019}

\begin{document}
\label{firstpage}
\pagerange{\pageref{firstpage}--\pageref{lastpage}}
\maketitle

\begin{abstract}
CTB~87 (G74.9+1.2) is an evolved supernova remnant (SNR) which hosts a peculiar pulsar wind nebula (PWN). The X-ray peak is offset from that observed in radio and lies towards the edge of the radio nebula. The putative pulsar, CXOU~J201609.2+371110, was first resolved with \textit{Chandra}
and is surrounded by a compact and a more extended X-ray nebula.
Here we use a deep {\textit{XMM-Newton}} observation to examine the morphology and evolutionary stage of the PWN and to search for thermal emission expected from a supernova shell or reverse shock interaction with supernova ejecta.
We do not find evidence of thermal X-ray emission from the SNR and place an upper limit on the electron density of 0.05~cm$^{-3}$ for a plasma temperature $kT\sim 0.8$ keV. The morphology and spectral properties are consistent with a $\sim$20~kyr-old relic PWN expanding into a stellar wind-blown bubble.
We also present the first X-ray spectral index map from the PWN and show that we can reproduce its morphology by means of 2D axisymmetric relativistic hydrodynamical simulations. 
\end{abstract}


\section{Introduction}
Rapidly rotating neutron stars drive magnetized winds of relativistic particles which, upon confinement by the surrounding supernova ejecta or the interstellar medium (ISM), form pulsar wind nebulae (PWNe). 
The emission from radio through the X-ray band is synchrotron radiation from high-energy particles moving in a strong magnetic field. CTB 87 is a plerionic supernova remnant (SNR) which shows a centrally filled morphology with no evidence of a shell at any wavelength. It has a radio size of 8$^{\prime}$ $\times$ 6$^{\prime}$ with a central flux density of 9 Jy at 1 GHz (\citealt{GreenCatalogue}) and a spectral break at 11 GHz (\citealt{6cmPolarization-spBreak}). HI data from the Canadian Galactic Plane Survey (\citealt{Kothes-HI-distance}) provides a distance estimate of 6.1~kpc. Recently \cite{Liu2018} found a superbubble, $\sim 37'$ in radius surrounding the SNR using HI 21~cm, WISE mid-IR, and optical extinction data. A previous \textit{Chandra} observation
\noindent(\citealt{Matheson2013}) identified a point source, CXOU J201609.2+371110, as the putative pulsar, and found an offset of 
$\sim 100^{\prime\prime}$ between the X-ray and radio peaks. Paired with the cometary morphology of the X-ray emission, this suggested the radio is 
either an evolved PWN due to supersonic motion of the neutron star, or a relic nebula resulting from interaction with the reverse shock. Observations in $\gamma$-rays with MILAGRO revealed an unresolved source, MGRO J2019+37 (\citealt{Abdo07}). This was later resolved by VERITAS as two sources, of which VER~J2016+371 is spatially coincident with CTB~87 (\citealt{Aliu14}). \textit{Fermi}-LAT found a source nearby to this VERITAS source in each of the 3 FGL catalogs (\citealt{Abdo10,Nolan12,Acero15}), the high energy FHL catalogs (\citealt{Ackermann13,Ackermann16};\linebreak \noindent\cite{Ajello17}), and the supernova remnant catalog (\citealt{Acero16}). The relation of this source to CTB~87 is unclear. The \textit{Fermi} source position is slightly different depending on the energy band used. \cite{Abeysekara18} argues this is the result of two unresolved high energy $\gamma$-ray sources, an idea supported by the evidence of variability in the low-energy FGL catalog analysis with reduced evidence in the FHL analysis. When the \textit{Fermi} emission is modeled as two point sources, a single power-law is required to fit the CTB~87 point source and the TeV VERITAS emission.

This paper is aimed at addressing the nature of the PWN through a deep \textit{XMM-Newton} study combined with numerical simulations.
In particular, we aim to address the cometary morphology of the PWN, constrain its spectral properties and search for (the so far missing) thermal X-ray emission expected from a supernova shell or reverse shock interaction with supernova ejecta. This is particularly important for addressing the nature of the growing class of shell-less PWNe which currently constitute 8--15\% of the SNR population \citep{2012AdSpR..49.1313F}\footnote{http://snrcat.physics.umanitoba.ca}.
Section 2 describes the observations. In Sections 3 and 4, we present our imaging and spectroscopy study results, including the first spectral index map of the PWN and the search for thermal X-ray emission from the SNR. In Section 5, we discuss these results in light of the evolutionary scenario for CTB~87 and the properties of its putative pulsar.
Section 6 presents our numerical simulations aimed at reproducing the morphology of the compact X-ray nebula and spectral index map. Finally, we summarize our results and conclusions in Section 7.

\section{Observations}

CTB 87 (G74.9+1.2) was observed with \textit{XMM-Newton} (ObsID 0744640101) for 125 ks starting 14 December 2014. The European Photon Imaging Camera (EPIC) detectors; MOS1, MOS2, and pn were operated in full frame imaging modes with the medium filter. Data were processed with the Science Analysis System (SAS) Version 16.1.0. Filtering for good time intervals, bad pixels, and out of time events was performed with \textit{emchain/epchain} for MOS and pn, respectively. Data products were then more strictly filtered for flaring with \textit{evselect} and \textit{tabgtigen} based on the appearance of the light curves. The effective exposure following filtering was 100~ks for MOS1, 105~ks for MOS2, and 75.4~ks for pn. Spectra were extracted separately from each detector with \textit{evselect} and fit simultaneously using the X-ray spectral fitting software XSPEC v12.9.1 (\citealt{XSPEC}) over the  $0.3-10$~keV energy range.

\section{Imaging}
\begin{figure}
\includegraphics[width=\columnwidth]{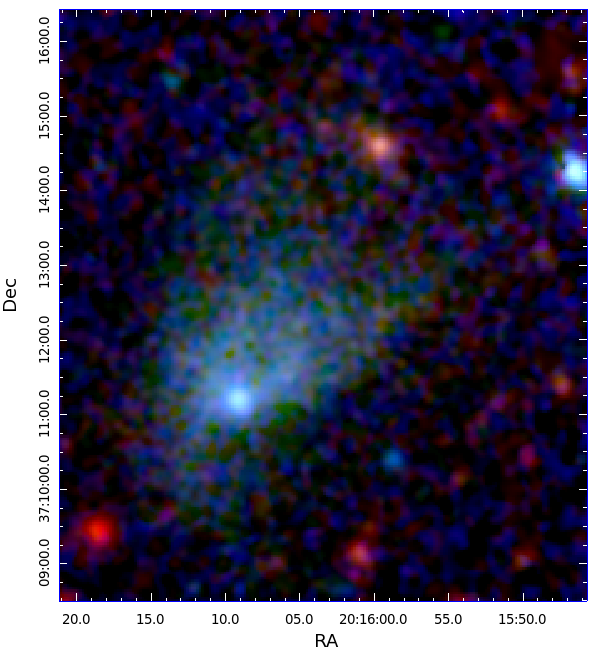}
\caption{Merged 
MOS1, MOS2 and pn RGB image adaptively smoothed to a signal-to-noise ratio of 4. The energy bands are Red: 0.3-1.5~keV, Green: 1.5-5~keV, Blue: 5-10~keV.}
\label{FIG:RGB}
\end{figure}

\cite{Matheson2013} used the superb angular resolution of the \textit{Chandra} X-ray Observatory to identify a point source as the putative pulsar, and arcs at the edge of the X-ray nebula which extend towards the radio peak. We recognize the pulsar as the bright source towards the centre of the PWN (Figure \ref{FIG:RGB}), and observe a cometary morphology with the PWN appearing brighter to the north-west of the pulsar. The arcs seen with \textit{Chandra}, however, are not readily apparent. An abundance of faint nebular emission not seen before appear in the \textit{XMM-Newton} data extending to the south-east, with the south-west edge lying along a chip gap in the pn detector.

\section{Spectroscopic Analysis}

Here we present a spatially resolved spectroscopic analysis of the different components of CTB~87.
Our spectroscopic study is targeted to constrain the parameters of the pulsar candidate and associated nebula (\S4.1 and \S4.2), map the spectral index across the SNR (\S4.3 and \S4.4), as well as search for any thermal X-ray emission that could be associated with the SNR (\S4.5).
\textit{XMM-Newton} has the advantage (over \textit{Chandra}) of its sensitivity to low-surface brightness emission.

\subsection{Compact Object}
\begin{figure}
    \centering
    \includegraphics[angle=90,width=\columnwidth]{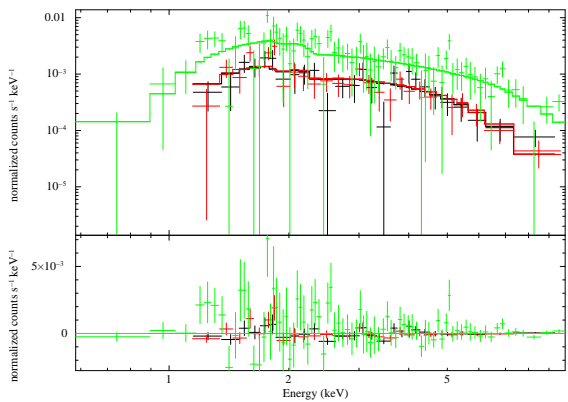}
    \caption{Pulsar spectrum and residuals from the MOS (black, red) and pn (green) detectors plotted with the best fit model.}
    \label{fig:CTB87-PulsarSpectrum}
\end{figure}

Spectra from the putative pulsar were extracted from a circle centred  at $\alpha (2000) = 20^h19^m9^s.2$, $\delta = +37^o11'10"$. Background spectra were extracted from a surrounding annulus. Due to the differing PSF of the detectors, the extraction radii are 8$^{\prime\prime}$ for MOS1 and MOS2 and 11$^{\prime\prime}$ for pn, with backgrounds spanning 10$^{\prime\prime}$--13$^{\prime\prime}$ and 12$^{\prime\prime}$--15$^{\prime\prime}$, respectively. The spectra were fit with an absorbed power-law with the absorption given by the Tuebingen-Boulder ISM absorption model (tbabs) with the abundances from \cite{Tbabs}. The column density was frozen to $2.249\times 10^{22}$~cm$^{-2}$, the best fit value found from fitting the diffuse nebula emission (see section \ref{Sec:DiffuseNebula}). The best fit yields a photon index $\Gamma = 1.62 \, (1.48-1.77)$ and an absorbed flux of $F_{abs,0.3-10~keV} = 7.56 \times 10^{-13} \,$ erg  cm$^{-2} \, $s$^{-1}$ where the parameters are reported with 90\% confidence intervals. The spectrum is shown in Figure \ref{fig:CTB87-PulsarSpectrum}.  

\subsection{Diffuse Nebula}
\label{Sec:DiffuseNebula}
\begin{figure}
\includegraphics[width=\columnwidth]{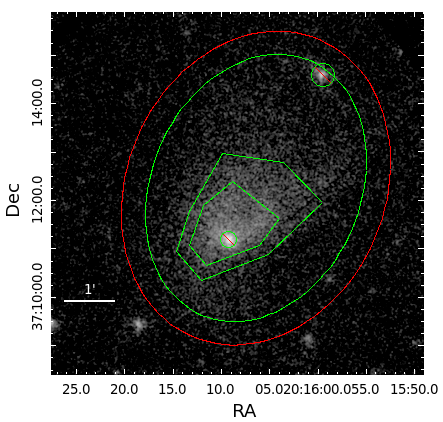}
\caption{Regions used to examine the diffuse nebular emission. The green ellipse defines the total diffuse region, while the red bounds the annulus used for background subtraction. The two polygons bound the inner and mid regions (Table \ref{TAB:Diffuse}).}\label{FIG:DiffuseRegions}
\end{figure}

To examine the properties of the extended nebular emission we extract spectra from the outer green ellipse ( $171^{\prime\prime} \times 129^{\prime\prime}$) in Figure \ref{FIG:DiffuseRegions} with a 10$^{\prime\prime}$-radius circle containing the pulsar and the unrelated point source to the north excluded. Background was extracted from a surrounding elliptical annulus ($200^{\prime\prime} \times 160^{\prime\prime}$ shown in red in Fig.~\ref{FIG:DiffuseRegions}). The best fit result was found with an absorbed power-law  model with a column density 
2.25 (2.10 -- 2.40) $\times~10^{22}$~cm$^{-2}$, photon index $\Gamma = $ 1.91 (1.84 -- 1.99).
This column density was then frozen for the subsequent fits to the inner regions (Table \ref{TAB:Diffuse}).

\begin{table*}
\begin{tabular}{l c c c c c}
\hline \hline
	&	Total	&	Outer	&	Mid	&	Inner	& Pulsar	\\\hline
$N_{H}$ $10^{22}$ $^{a}$ &	2.25 (2.10 - 2.40)	&	2.25	&	2.25	&	2.25	& 2.25\\
Photon index, $\Gamma$	&	1.91 (1.84 - 1.99)	&	2.18 (2.09 - 2.27)	&	1.93 (1.88 - 1.98)	&	1.76 (1.73 - 1.80)	& 1.62 (1.48 - 1.77)\\
Norm $10^{-4}$ $^{b}$	&	5.61 (5.05 - 6.26)	&	2.62 (2.38 - 2.87)	&	1.45 (1.37 - 1.53)	&	1.65 (1.58 - 1.72)	& 0.17 (0.15 - 0.20)\\
$\chi^{2}(\nu)$	&	1.080 (1039)	&	1.112 (675)	&	1.055 (612)	&	0.931 (664)	& 0.998 (133)\\
Flux $10^{-13}$ $^{c}$	&	16.01 (15.71 - 16.23)	&	5.07 (4.91 - 5.24)	&	4.01 (3.92 - 4.10)	&	5.85 (5.77 - 5.94)	& 0.76 (0.72 - 0.78)\\
\hline
\multicolumn{6}{l}{Confidence ranges are 90\%. Models were fit over the range 0.3--10 keV}\\
\multicolumn{6}{l}{$^{a}$ cm$^{-2}$}\\
\multicolumn{6}{l}{$^{b}$ photons keV$^{-1}$ cm$^{-2}$ s$^{-1}$}\\
\multicolumn{6}{l}{$^{c}$ erg cm$^{-2}$ s$^{-1}$, 0.3-10 keV observed.}\\
\end{tabular}

\centering\caption{Absorbed power-law model fits to the regions shown in Fig.~\ref{FIG:DiffuseRegions}.}\label{TAB:Diffuse}
\end{table*}

\subsection{Radial Profile}
\label{Sec:RadialProfile}
\begin{figure}
\includegraphics[width=\columnwidth]{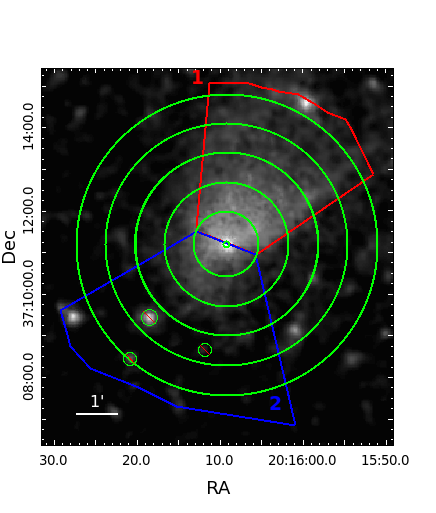}
\caption{Regions used to extract spectra for the radial profile. Point sources (shown in circles) were removed before fitting the PWN's spectra.}\label{Fig:PieRegions}
\end{figure}

\begin{figure}
\includegraphics[width=\columnwidth]{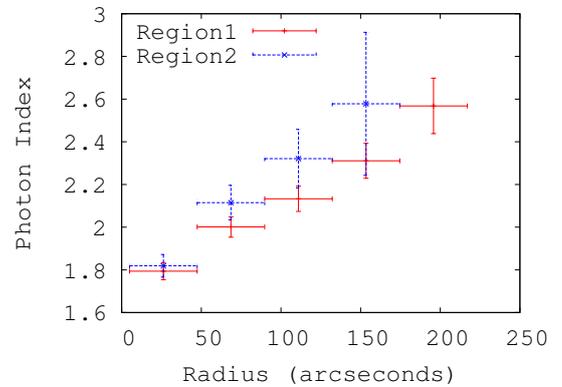}
\caption{Results of radial profile analysis showing the photon index at increasing distance from the pulsar for the regions shown in Figure \ref{Fig:PieRegions}. Region 1 follows the cometary morphology back towards the radio peak. Region 2 covers the emission ahead of the proposed bow shock. We do not display the outermost section of region 2 as it does not contain sufficient emission to fit a spectrum.}\label{FIG:RadialProfile}
\end{figure}

One of the main questions surrounding CTB~87 is the nature of its morphology. The previous \textit{Chandra} study by\\ \noindent\cite{Matheson2013} identified a bow-shock or cometary appearance of the X-ray PWN and favoured a reverse shock interaction model rather than one of ram-pressure confinement due to an inferred high pulsar velocity. We examine spectra from opposite directions of the pulsar candidate to search for evidence of an interaction. The radial profile regions (Figure \ref{Fig:PieRegions}) are selected by generating 5 concentric rings centred on the pulsar from $5^{\prime\prime} - 217^{\prime\prime}$, with the angular extent given by the polygon which bounds the cometary morphology to the north west, and a symmetric polygon to the south east.
As the reverse shock encounters the PWN, it is predicted to compress the local magnetic field, leading to a more rapid synchrotron burn off, and softening of the spectral index (e.g., \cite{KC84}). 
 Comparing the profiles (Figure \ref{FIG:RadialProfile}), we find that the spectral indices are consistent within error, suggesting that if a reverse shock interaction has occurred, enough time has elapsed for the pulsar wind to return the nebula to equilibrium. We note that there is an apparent trend of softer indices for the southern (blue) region shown in Figure \ref{Fig:PieRegions} out to a 150$^{\prime\prime}$-radius, however the errorbars are larger for this region given the lower surface brightness. A deeper observation is needed to study the southern outflow, particularly visible in Figure \ref{FIG:DiffuseRegions}, to constrain its spectral index and its variation.

\subsection{Photon Index Map}
\begin{figure}
\includegraphics[width=\columnwidth]{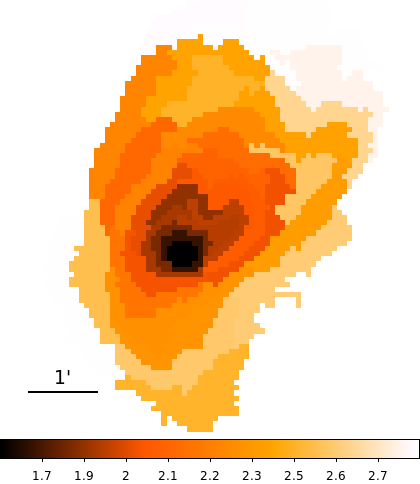}
\caption{Adaptively binned regions covering the PWN, each fit with an absorbed power-law and coloured by photon-index. The cometary morphology is apparent with trailing arms. There is also extended emission to the south.}
\label{FIG:SpectralMap}
\end{figure}

The apparent pure non-thermal nature of CTB~87 makes traditional RGB images appear simplistic with the only observable being a slight softening of the bulk emission with distance from the pulsar. To search for fine detail and hidden features in the non-thermal emission, we generate a photon index map. Using the adaptive binning software \textit{Contbin} (\citealt{Contbin}) we generate regions which meet a signal limit (chosen such that each contains at least 400 counts by the MOS detectors), extract spectra and fit with an absorbed power-law. The absorption was frozen to the best fit value from the total nebula region. The spectral map (Figure \ref{FIG:SpectralMap}) shows steepening with distance from the central pulsar, along with trailing arms stretching to the north-west towards the radio peak.

\subsection{Search for Thermal Emission}
Our RGB image (Figure \ref{FIG:RGB}) shows no clear signs of excess soft X-ray emission attributable to shocked ISM or ejecta. We therefore take two approaches for our search for thermal X-ray emission from CTB~87. First, we extract spectra from an elliptical annulus region surrounding the X-ray PWN and  fit it with an absorbed thermal model ($APEC$ in XSPEC) which refers to the `Astrophysical Plasma Emission Code' (\citealt{Smith01}) and describes collisionally ionized diffuse gas calculated from the AtomDB atomic database. This gives an upper limit on the observed thermal flux from a surrounding shock-heated medium of $1.85\times 10^{-13}$~erg cm$^{-2}$ s$^{-1}$. If the cometary morphology is due to the interaction with a reverse shock, we expect the dominant source of thermal emission to be along the bow-shock like feature. Our second approach to place limits on the thermal emission is to freeze the best fit power-law parameters from the total diffuse region and add a thermal APEC component until the model exceeds the data by $2\sigma$. This provides an observed flux limit of $8.687\times 10^{-15}$ erg~cm$^{-2}$~s$^{-1}$ for a temperature of kT = 0.82 keV, the best fit temperature from the surrounding region. We later investigate the density implied by a range of temperatures (Section \ref{Sec:ThermalLimits}).

\begin{figure}
\includegraphics[width=\columnwidth]{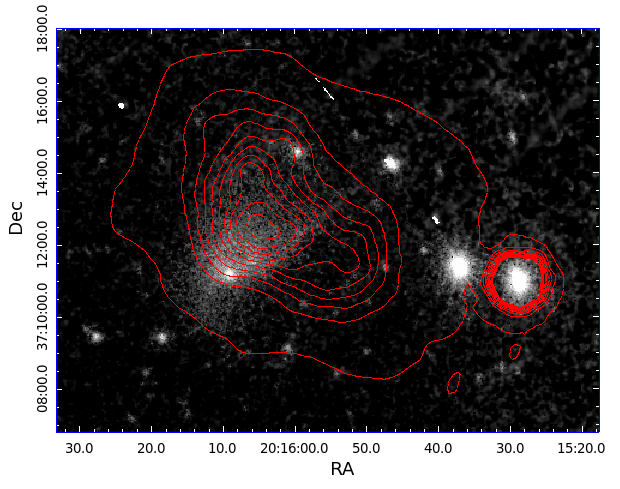}
\caption{Merged MOS1, MOS2, and pn image overlayed with the radio contours from the Dominion Radio Astrophysical Observatory as part of the Canadian Galactic Plane Survey (\protect\cite{Radio-Contours}).}\label{FIG:Contours-Image}
\end{figure}

\section{Discussion}
\subsection{Putative Pulsar Properties}
To date no pulsations have been observed from CTB~87, however we may estimate the properties of the neutron star using its X-ray flux and the empirical results of \cite{PulsarProperties2008}. The unabsorbed putative pulsar's luminosity in the 2--10 keV band is $3.5\times 10^{32} D^2_{6.1}$ erg s$^{-1}$. The spin-down energy loss $(\dot{E})$ is then given by $L_\mathrm{X,PSR} = 10^{-0.8}\dot{E}^{0.92}$ leading to $\dot{E} = 1.76\times 10^{36}$ erg s$^{-1}$. Alternatively an estimate can be made from the unabsorbed PWN luminosity in the same band ($L = 7.4\times 10^{33}D^2_{6.1}$ erg s$^{-1}$) and the spin-down energy loss may be predicted from $L_\mathrm{X,PWN}=10^{-19.6}\dot{E}^{1.45}$ which gives $\dot{E}=7.5\times 10^{36}$ erg s$^{-1}$.

\cite{PulsarProperties2008} also provide empirical results for the characteristic age ($\tau$) of a pulsar given either the pulsar or PWN luminosity. These are, respectively, $L_\mathrm{x,PSR}=10^{38.9}\tau^{-1.4}$ yr, and $L_\mathrm{x,PWN}=10^{42.4}\tau^{-2.1}$ yr, leading to ages of $\sim$34.5~kyr and 11.5~kyr respectively. This is comparable to the 5--28 kyr estimate from \cite{Matheson2013} which was based on typical neutron star kick velocities and the assumption of the radio emission belonging to a relic nebula formed at the birth location of the pulsar.

Combining this age with the predicted spin-down energy loss allows an estimate of the pulsar properties through the spin-down energy loss and pulsar braking index relations (described in, e.g., \cite{GaenslerSlane2006}). The period ($P$) is given by $P=((4\pi^2I)/((n-1)\dot{E}\tau))^{1/2}$, the period derivative by $\dot{P}=P/((n-1)\tau)$, and surface magnetic field by  $\text{B} = 3.2\times 10^{19}I^{1/2}_{45}R^{-3}_{10}(P\dot{P})^{1/2}$. Here $n$ is the braking index, $I_{45}$ is the moment of inertia of the neutron star in units of $10^{45} \text{g}\, \text{cm}^{2}$, and $R_{10}$ is the radius in units of 10 km.  Table \ref{tab:PSR-Characteristics} lists estimates of these values based on the two different calculated energy loss rates and an adopted average age of 20~kyr. These properties are within the range expected for rotation-powered pulsars with prominent PWNe. Previous searches in radio have failed to find pulsations down to a flux limit of $\sim$1~mJy (\cite{Gorham96,Lorimer98,Straal19}).
Future sensitive timing observations of CXOU~J201609.2+371110 in radio and X-ray are needed to pin down the properties of this pulsar candidate.

\begin{table}
    \centering
    \begin{tabular}{l|c|c}\hline\hline
        $\dot{E}$ & $1.76\times 10^{36}$ erg s$^{-1}$ & $7.5\times 10^{36}$ erg s$^{-1}$ \\\hline
        $P$ $^a$ & 0.13 s & 0.065 s \\
        $\dot{P}$ $^b$  & $1.06 \times 10^{-13}$ & $0.51\times 10^{-13}$  \\
        $B$ $^c$   & $1.8\times 10^{12}$~G  & $3.8\times 10^{12}$~G  \\\hline
         \multicolumn{3}{l}{$^a$ $I^{1/2}_{45}\dot{E}^{-1/2}_{36}t^{-1/2}_{20}$  \, $^b$ $I^{1/2}_{45}\dot{E}^{-1/2}_{36}t^{-3/2}_{20}$ \, $^c$ $I_{45}\dot{E}^{-1/2}_{36}R^{-3}_{10}t^{-1}_{20}$}\\
    \end{tabular}
    \caption{Pulsar properties based on the spin-down energy loss derived from the X-ray luminosity of the pulsar (left) and PWN (right), and a characteristic age of 20 kyr.}
    \label{tab:PSR-Characteristics}
\end{table}

\subsection{Limits on Ambient Density and Faint Thermal Emission}\label{Sec:ThermalLimits}
\begin{figure}
\includegraphics[width=\columnwidth]{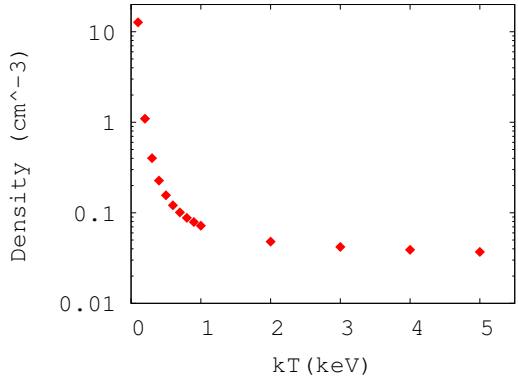}
\caption{Upper limit on the ambient density derived from adding a thermal bremsstrahlung component with fixed temperature.}\label{Fig:DensityLimits}
\end{figure}

No thermal X-ray emission nor limb-brightening has been detected from CTB~87.
The lack of any observed SNR shell suggests expansion into a low-density medium. To examine the level of density down to which our observations are sensitive we set an upper limit on the density of emitting electrons by estimating 
the upper limit on any thermal contribution to the total diffuse emission. The
normalization of an additional thermal component in the diffuse region has an upper limit of $\frac{10^{-14}}{4\pi D^{2}}\int n_{e}n_{H}dV= 5.67\times 10^{-5}$cm$^{-5}$ (for $kT$=0.82 keV). This implies that $\int n_{e}n_{H}dV \sim fn_{e}n_{H}V = 10^{14}(4\pi D^{2})(5.67\times 10^{-5}cm^{-5})$, where $f$ is the volume filling factor, $n_{e}$ is the electron density, $n_{H}\sim n_{e}/1.2$. The total volume $V$ is assumed to be an ellipsoid with semi-axis of a=b=171$^{\prime\prime}$, and c=129$^{\prime\prime}$. This volume projected is the ellipse shown in Figure \ref{FIG:DiffuseRegions}. The electron density upper limit is then $n_{e}<0.05f^{-1/2}D^{-1/2}_{6.1}$ cm$^{-3}$. To understand how our assumption of the temperature of the emitting electrons affects the density we freeze the temperature parameter and derive the corresponding upper limit for the electron density. The results are shown in Figure \ref{Fig:DensityLimits}. The high temperature and low inferred density are consistent with the picture of expansion into a stellar wind blown bubble 
(\citealt{Matheson2013,Liu2018}, Kothes et al. in prep.)

According to the radio study (Kothes et al., in prep), the shell-type SNR has a radius which would place it outside the \textit{Chandra} and \textit{XMM-Newton} observations in any direction. In addition,
any thermal emission from those parts of the shell that is moving towards us or away from us would be too faint and diffuse to be detected. Because of the large size, its thermal emission would have been subtracted with the estimated background.

The other thermal component that could be detected is the emission from the supernova ejecta. The radio study (Kothes et al., in prep) places the blast wave of the supernova explosion, which is still expanding freely inside the HI cavity discovered by \citet{Liu2018}, at a radius larger than 30~pc. For a maximum ejecta mass of 20~M$_\odot$, evenly distributed inside the 30~pc, we would find an average electron density of about n$_e = 0.006$~cm$^{-3}$ assuming it consists of 90~\% hydrogen and 10~\% helium and the material is fully ionized. This is obviously an upper limit since the ejecta should also contain heavier elements, in which case we would get a lower number of electrons per mass unit. This is well below our detection threshold for any reasonable temperature.

CTB87 is likely interacting with molecular material discovered by \citet{Kothes-HI-distance} and later confirmed by \citet{Liu2018}. In the direction of this molecular cloud complex a candidate for a radio shell was detected with a radius of about 13~pc (Kothes et al., in prep). If we distribute ejecta of 20~M$_\odot$ evenly inside this shell, we get an average electron density of n$_e = 0.075$~cm$^{-3}$, again, fully ionized, with 90\% hydrogen and 10\% helium. This is slightly above our detection threshold, but this is of course again an upper limit since we do not take heavier elements into account. Below 15~M$_\odot$ the thermal emission from the ejecta would definitely be undetectable.

\subsection{Morphology}
The overall appearance of CTB~87 is similar to G327.1--1.1 (\citealt{Temim15}), and MSH 15--56 (\citealt{Yatsu13}) which display a cometary morphology. However, both are embedded in SNR thermal shells which we do not detect in our deep \textit{XMM-Newton} observation. G327.1--1.1 displays prongs of non-thermal emission extending ahead of the bow-shock which has been proposed to be a limb-brightened cone and is similar to the emission we see extending to the south-east of the putative pulsar. The lack of thermal emission in CTB~87 may be the result of expansion into a low-density medium. Emission from high-temperature, low-density plasma mimics a power-law and the high column density masks any faint soft X-rays.

The brightest radio emission forms a kidney shape with the elbow pointing towards the X-ray peak (Figure \ref{FIG:Contours-Image}). The X-ray emission likewise displays a cometary morphology with trailing arms which point back towards the radio centre. There are two main accepted possibilities for the source of this morphology. The neutron star may be moving supersonically through the ambient medium, leaving a relic radio nebula behind, and driving a bow shock. Alternatively, the morphology may be due to an interaction with the reverse shock of the SNR shell. In order to drive a bow shock, the ram pressure due to the pulsar's proper motion ($\rho_{0} v^{2}_\mathrm{PSR}$, where $\rho_{0}$ is the density of the ambient medium) must exceed the nebular pressure. \cite{Matheson2013} argued against this due to the unlikely high pulsar velocity needed. To drive a bow-shock, the pulsar must satisfy $v_\mathrm{PSR}\geq 325 n_{0}^{2/17}E_{51}^{1/17}$ km s$^{-1}$ (\citealt{VanDerSwaluw04}, where $E_{51}$ is the explosion energy of the supernova and $n_{0}$ is the ambient density. The required velocity for the lower limit on density given a hot thermal component with $kT \gtrsim 1$~keV,  $\rho \approx 0.05\text{cm}^{-2}$ (Figure \ref{Fig:DensityLimits}) is 218 km s$^{-1}$. Assuming the radio emission is from a relic PWN, the offset between the radio and X-ray peak locations of $\sim 140''$ (see Fig.~6) leads to a maximum age of 19~kyr for motion in the plane of the sky, and older for an arbitrary inclination angle.  A bow shock origin also brings into question the nature of the faint extended emission preceding the pulsar (to the south-east/south, see Figs. \ref{FIG:RGB} and \ref{FIG:DiffuseRegions}) which is much more pronounced in this \textit{XMM-Newton} observation.

Arguing for the reverse shock scenario proves to be difficult due to the absence of observed thermal X-ray emission. The very low density derived in Section \ref{Sec:ThermalLimits} is consistent with expansion into low-density ejecta expanding into a stellar wind bubble, in agreement with the results of \cite{Liu2018} and Kothes et al. 2019 (in preparation). A reverse shock interaction may however explain the the steep radio spectrum observed from the central kidney-shaped component. This component is embedded in a large diffuse hard-spectrum region which may represent the unperturbed PWN (Kothes et al., in preparation).

\section{Numerical simulations}
\label{sec:NSim}
The morphology of CTB~87 is reproduced here by means of 2D axisymmetric relativistic hydrodynamical (HD) simulations based on the numerical code PLUTO \citep{Mignone:2007a}. 
The aim of this numerical study is to verify how the pulsar parameters inferred from observations match the shape of the compact X-ray nebula, constraining  the possible geometry of the system, in particular focusing on reproducing the X-ray spectral index map.

The simulation has a cylindrical grid, with a domain range $z\in[-110,50]$ ly and $r\in[0,80]$ ly, corresponding to a $50\times25$ pc$^2$ box, and a resolution at the base level of $272\times544$ cells.
Increased resolution near to the pulsar location and at the shock is obtained by imposing 5 refinement levels thanks to Adaptive Mesh Refinement (AMR) facilities \citep{Mignone:2012}, achieving a maximum resolution of $8704\times17408$ cells. 
The simulation employs an HLL Riemann solver, a second order Runge-Kutta time integrator and a Van Leer Limiter. The equation of state is set to be ideal, with the appropriate adiabatic index for describing a relativistic plasma ($\gamma_A=4/3$), and the relativistic nature of the unshocked pulsar wind is ensured by imposing a wind Lorentz factor of 10. 

The simulation has the pulsar located at the centre of the grid, and a wind is injected within a small region in its surroundings (details of the wind modeling can be found in \citealt{Olmi:2015}). The interaction of the pulsar wind with the surrounding ejecta of the supernova explosion generates the observed PWN. 
The variation of the pulsar energy output with time is given by $L(t) = \dot{E}_0/(1+t/\tau_0)^{(n+1)/(n-1)}$, with $\dot{E}_0$ the initial spin-down energy, $n$ the braking index and $\tau_0$ the spin-down age of the system. 
Since the pulsar powering the PWN has not yet been identified, the exact properties of the pulsar are not known. Moreover there are no clear indications about the location of the supernova remnant shell.

We have thus considered average values of the parameters as inferred from the spectral study of well known systems (see for example \citealt{Torres2014})
 imposing $E_\mathrm{SN}=10^{51}$ erg, $M_\mathrm{ej}=12 M_{\odot}$ and $n=3$.

The initial spin-down energy $\dot{E}_0= 5\times10^{38}$ erg s$^{-1}$  and spin-down time $\tau_0= 2150$ yr
have been chosen in order to get a spin-down luminosity of $\dot{E} \simeq 4.8 \times 10^{36}$ erg s$^{-1}$ at the age of $\sim$20 kyr, corresponding to a rough  average of the values  estimated from the range of X-ray luminosities (see Table 2).
Those values are chosen in order to obtain a qualitative matching between the simulation and observations of the compact X-ray nebula. 
In any case different choices for the parameters are not expected to change the dynamical evolution of the system, ensuring that the characteristic time and radius of the system, as defined in \citet{Truelove&McKee1999}, are kept fixed. For example this means that the same results can be obtained considering: $E_\mathrm{SN}= 5.2\times 10^{50}$ erg, $M_\mathrm{ej}= 8.1M_{\odot}$, $\dot{E}_0= 9.6\times10^{38}$ erg s$^{-1}$ and $\tau_0= 2000$ yr; 
 or $E_\mathrm{SN}=10^{51}$ erg, $M_\mathrm{ej}= 20 M_{\odot}$, $\dot{E}_0= 3.3\times10^{38}$ erg s$^{-1}$ and $\tau_0= 3060$ yr.  

\begin{figure}
	\centering
	\includegraphics[width=.45\textwidth]{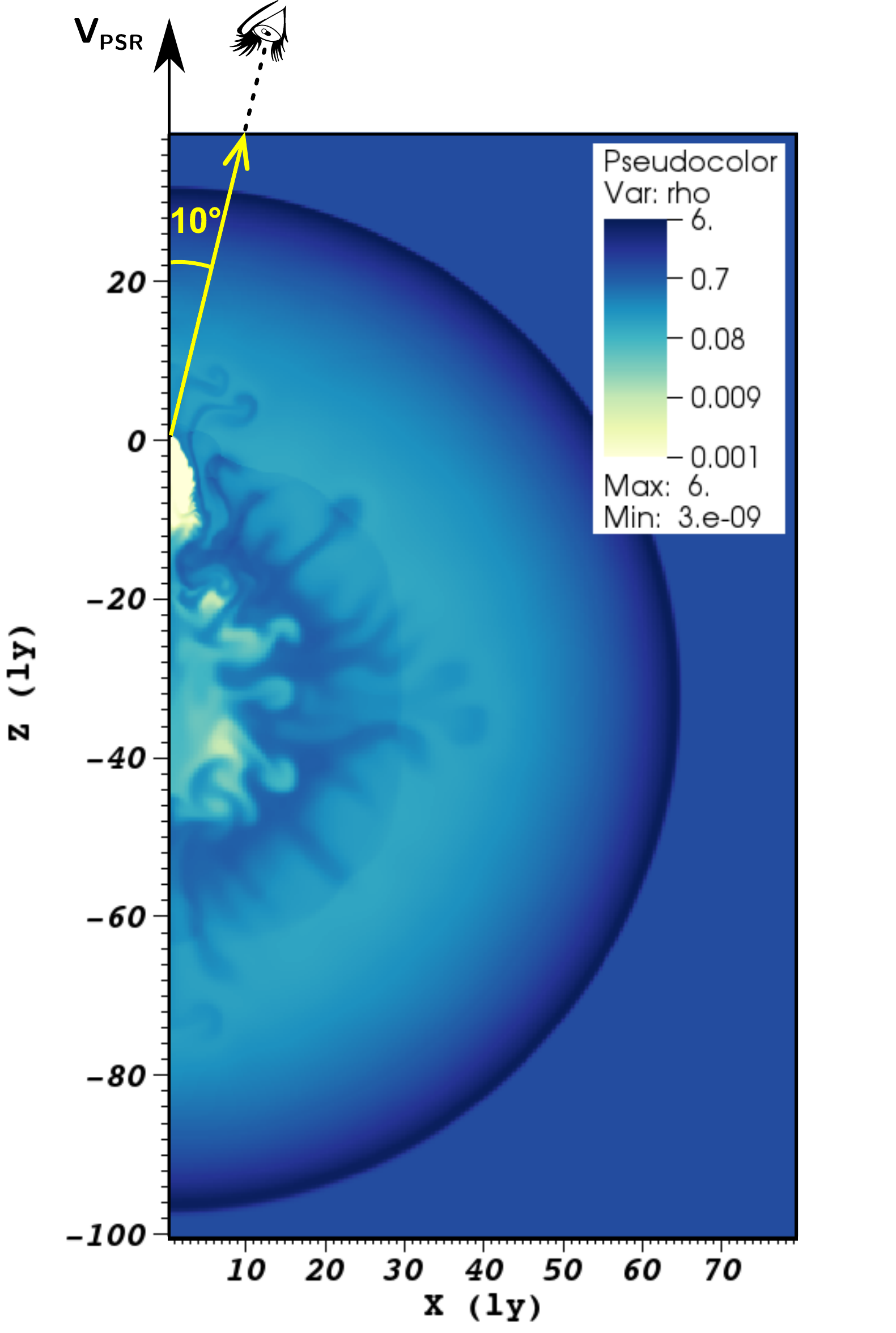}
	\caption{Half of a planar map of the density in units of $10^{-24}$ particles/cm$^{3}$, with logarithmic scale.  The yellow line indicates the line of sight inclination with respect to the pulsar direction of motion (the $z$ axis). }
	\label{fig:sim_incl}
\end{figure}

Based on the analysis of the X-ray observations presented in \citet{Matheson2013}, we suggest that the pulsar kick velocity is inclined by $\alpha=100\degr$ with respect to the plane of the sky, i.e., with an inclination of $10\degr$ with respect to the line of sight, and rotated by $\theta\sim30\degr$ towards the NW direction. The inclination of the line of sight is shown in Fig.~\ref{fig:sim_incl}, superposed on a density map, in logarithmic scale. The figure shows that the ambient density is rather low, of the order of $\rho_0 \sim 0.08-0.5 \times 10^{-24}$ g/cm$^{-3}$, in agreement with estimates and compatible with typical values of the ejecta density.
According to the described geometry we impose a kick velocity of $\mathrm{v}_\mathrm{PSR}\sim 400$ km/s, in order to match the expected lower limit of the planar velocity ($\sim 220$ km/s). 
The system is then evolved up to a final age of $t_f=20000$ yr, in agreement with the prediction from the analysis of the X-rays and radio emission. At this age the PWN has an elongated morphology due to its proper motion with respect to the centre of the explosion and it has already interacted with the SNR reverse shock.

Let us stress here that, from the modelling point of view, the reverse shock scenario and the bow shock one are similar. The deformation of the PWN morphology produced by an incoming reverse shock moving through a non-uniform medium would produce dynamically the same effect of the proper motion of the pulsar in the ambient (uniform) medium, as considered in the present case. Only the combined case of the proper motion and the anisotropy of the ambient medium could lead to a more complex dynamics (\citealt{Kolb:2017}).
Our predictions remain thus valid in the light of our qualitative analysis for both scenarios.

Based on our initialisation values, the radius of the remnant at the present age $t_f=20$ kyr is
$R_\mathrm{SNR} \sim 6 \times 10^{19}$ cm or $\sim 20$ pc (or $\sim 60~ly$ as can be seen in Fig.~\ref{fig:sim_incl}). The pulsar is then located at about $0.5 R_\mathrm{SNR}$, which is close to the transonic limit of $\sim 0.7 R_\mathrm{SNR} $. The radius of the termination shock ($r_\mathrm{TS}$) can be then roughly estimated from the expression of the typical bow shock  stand-off distance, defined as $d_0=[\dot{E}/(4\pi c \rho_{0}\mathrm{v}^2_\mathrm{PSR})]^{1/2}$ based on balance of the wind momentum flux and the ram pressure of the ambient medium with local density $\rho_0$: $r_\mathrm{TS}\sim d_0 = 0.06$ pc. 
This is also in agreement with the direct estimation from dynamic maps, from which the termination shock is $r_\mathrm{TS}\sim 0.1$ pc. These values are both  roughly  consistent with the $\sim 0.15$ pc estimation obtained by \citet{Matheson2013} based on the distance of the X-ray inner arc from the putative pulsar position ($\sim 5^{\prime\prime}$) and the estimated distance of the source of 6.1 kpc. The difference between the estimate of $r_\mathrm{TS}$ based on dynamics and that derived from X-ray images is not significant, given the uncertainty in the identification of the termination shock and pulsar position in the X-ray maps. Notice moreover than the estimation from the bow shock scenario is strongly dependent on the value of the ambient density, which is not known with precision.

\begin{figure}
	\centering
	\includegraphics[width=.45\textwidth]{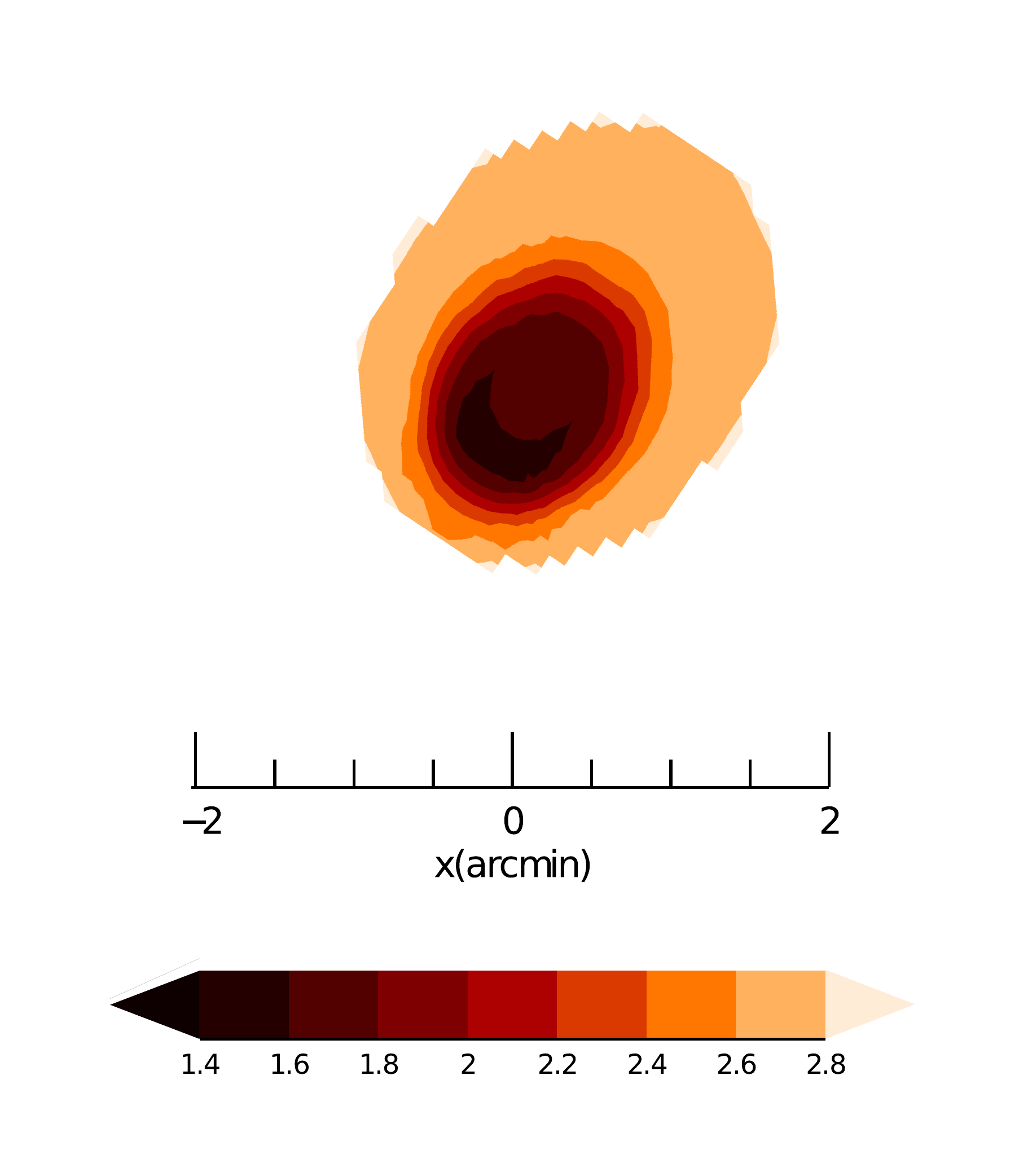}
	\caption{Map of the X-ray spectral index of CTB 87 in the 0.5-10 keV band. Value of the spectral index is given as contours of different colours expressed by the scale in the bottom of the map. Dimensions of the system are given in arcminutes. This map must be compared with Fig.~\ref{FIG:SpectralMap}, keeping in mind that the simulation can only reproduce the emission from the compact nebula, while the contribution to the spectral index map from the diffuse nebula is not considered here.
	\label{fig:sim_spINDX}}
\end{figure}

Fig.~\ref{fig:sim_spINDX} shows the synthetic X-ray spectral index map in the 0.5--10 keV range, to be compared with XMM-Newton observations. 
The synthetic map is built with the following procedure: the emission range is divided into 8 different, equally spaced, frequencies; and for each frequency the emissivity is computed with standard expressions for the synchrotron emission (as can be found for example in \citealt{Del-Zanna:2006}).
The value of the  magnetic field is assumed in the post-processing analysis by imposing that the magnetic pressure is a small fraction of the total pressure. The exact amount is  determined by the matching of the size of the emitting area with the one observed in the same energy range. This leads to an average magnetic field in the nebula of the order of $\sim 6 \,\mu$G, which is fully compatible with the estimation of a magnetic field of$ \sim 50\,\mu$G at the termination shock given in \citet{Matheson2013}, since strong magnetic dissipation is expected to take place as the flow moves from the inner to the outer nebula and a drop by a factor of 10 is not unrealistic (\citealt{Olmi:2016}, \citealt{Olmi:2019}).
Spectral index maps are then computed following the procedure given in \citet{Del-Zanna:2006}, and the final map is built as the combination of the single maps.

Fig.~\ref{fig:sim_spINDX} shows a rather impressive agreement with observations, both in terms of the expected shape and  of the shell-like distribution of values of the spectral index. The region surrounding the pulsar is well recognizable as the $\sim 1.5$ spectral index zone and is surrounded by shells of increasing value of the spectral index up to 2.8.
The major difference is in the diffuse emission in the farther region, which is also non-axisymmetric in the data and, of course, cannot be exactly reproduced with 2D axisymmetric simulations. 
In order to reproduce non axisymmetric features the system must be in fact simulated with full 3D geometry. On the other hand, this is computationally rather challenging. Such 3D simulations are very demanding in terms of numerical cost (core-hours and total duration of the simulation, as it is largely discussed in \citealt{Porth:2014, Olmi:2016}), since they require very high resolution at the grid centre, where the pulsar wind is injected. 
Moreover accounting for such a long evolution as required in the present case (10--20~kyr vs a few hundreds of years, the typical duration of 3D simulations) would require very long time-scales with the actual computational resources, making the 3D approach not feasible for our study.
On the other hand the 2D approach was largely shown to be able to match the main morphological properties of PWNe \citep{Blondin:2001, VanDerSwaluw04, Del-Zanna:2006, Olmi:2014, Olmi:2015}, while the HD approximation seems to be adequate in representing evolved systems, where the magnetic field is expected to be quite low and tangled in the bulk of the nebula, due to magnetic dissipation  (\citealt{Temim15}, \citealt{Kolb:2017}, \citealt*{Olmi:2019}).
Except for the obvious deformation from a perfect axisymmetric structure, we might thus expect that a 3D magneto-hydrodynamic representation would not change that much the results of the global morphology found with the present approach.

\section{Conclusions:}
 Using a 125~ks observation with \textit{XMM-Newton}, we do not find evidence of thermal X-ray emission, supporting the scenario for expansion into a stellar wind bubble. We present the first spectral index map of the PWN revealing high-resolution hard structures close to the pulsar candidate, and an overall steepening of the photon index away from it. We find low-surface brightness emission to the south of the pulsar candidate, which suggests diffusion of particles in that direction. The combined  X-ray and radio study of the compact nebula supports the relic PWN scenario resulting from interaction with the reverse shock of an evolved $\sim$20~kyr PWN. 

The geometry of the system is investigated by means of HD numerical simulations. Given the uncertainties in the parameters connected to the PSR and the SNR, we have considered a set of values compatible with average ones from the known PWN catalogue, according with the ranges identified from the multi-wavelength analysis of the system. An investigation of different sets of parameters would be computationally demanding and, in any case, the global morphology of the system is not expected to change if the parameters (i.e., $E_\mathrm{SN}$, $M_\mathrm{ej}$ and $\tau_0$) are varied keeping constant the characteristic time and radius as defined in \citet{Truelove&McKee1999}.
Moreover the primary intent here was to reproduce the morphology of the X-ray compact nebula, especially in terms of the spectral index map. These give us some indications on the possible geometry of the system on the plane on the sky, which would hopefully be better constrained in the future with new observations.
For instance variations of the inclination angle and PSR velocity can be easily accounted for in the same model, without affecting heavily the predictions.
The impressive agreement of the spectral index map with the \textit{XMM-Newton} observations supports the scenario of an evolved and elongated PWN moving towards the observer through low-density SNR material.

Timing and deeper observations extending beyond the radio nebula are needed to explore open questions about this system. In particular the spin properties of the pulsar candidate and a measurement of its proper motion will be crucial.
Furthermore, a deeper X-ray spectroscopic study extending to larger distances from the pulsar candidate may reveal weak thermal X-ray emission from this system.

\section*{Acknowledgements}
 This research was supported by the Natural Sciences and Engineering Research Council of Canada (NSERC) through the Discovery Grants and the Canada Research Chairs Programs (S.S.H.). B.G. acknowledges support from a University of Manitoba Graduate Scholarship.
E.A., N.B. and B.O. wish to acknowledge financial support from the ``Accordo attuativo ASI-INAF n. 2017-14-H.0, progetto: \textit{on the escape of cosmic rays and their impact on the background plasma}", the SKA-CTA-INAF and INAF-MAINSTREAM projects.
We acknowledge an earlier contribution by H. Matheson towards the XMM proposal.
We thank the referee for comments that helped improve the paper.
This study made use of NASA's Astrophysics Data System and ESA's \textit{XMM-Newton} facility.

\bibliographystyle{mnras}
\bibliography{ctb87}

\bsp	
\label{lastpage}

\footnotesize{

}

\end{document}